\documentclass[aps,pre,twocolumn,superscriptaddress,floatfix]{revtex4-1}

\usepackage{amsmath,amssymb}
\usepackage{bm}
\usepackage{mathptmx}
\usepackage{graphicx}
\usepackage{color}
\usepackage{hyperref}
\usepackage{float}
\usepackage{microtype}
\usepackage{xspace}
\usepackage{booktabs}
\usepackage{makecell}
\usepackage{nicefrac}
\usepackage[normalem]{ulem}
\usepackage{enumitem}

\newcommand{\PNG}{\textsc{png}\xspace}
\newcommand{\PNGs}{\textsc{png}s\xspace}
\newcommand{\SNN}{\textsc{snn}\xspace}

\newcommand{\STDP}{\textsc{stdp}\xspace}
\newcommand{\RP}{\textsc{rp}\xspace}
\newcommand{\RQA}{\textsc{rqa}\xspace}
\newcommand{\DET}{\mathrm{DET}}
\newcommand{\prw}{p_{\mathrm{rw}}}
\newcommand{\Wmax}{W_{\!\max}}

\begin{document}

\title{Topology-Dependent Emergence of Polychronous Neuronal\\
       Groups: A Recurrence-Plot Characterization}

\author{Lucas A.\,T.\,X. Carneiro}
\affiliation{%
  Complex Systems Modelling Graduate Program,
  School of Arts, Sciences and Humanities (EACH),
  University of S\~{a}o Paulo, 03828-000 S\~{a}o Paulo, SP, Brazil}

\author{Armand D. Jiofack}
\affiliation{%
  Department of Physics, Faculty of Philosophy, Sciences
  and Letters of Ribeir\~{a}o Preto (FFCLRP),
  University of S\~{a}o Paulo, 14040-900 Ribeir\~{a}o Preto, SP, Brazil}

\author{Fernando F.~Ferreira}
\email{ferfff@usp.br}
\affiliation{%
  Complex Systems Modelling Graduate Program,
  School of Arts, Sciences and Humanities (EACH),
  University of S\~{a}o Paulo, 03828-000 S\~{a}o Paulo, SP, Brazil}
\affiliation{%
  Department of Physics, Faculty of Philosophy, Sciences
  and Letters of Ribeir\~{a}o Preto (FFCLRP),
  University of S\~{a}o Paulo, 14040-900 Ribeir\~{a}o Preto, SP, Brazil}

\begin{abstract}
Polychronous Neuronal Groups (\PNGs) reproducible, time-locked
spatiotemporal firing cascades stabilised by Spike-Timing-Dependent
Plasticity (\STDP) and heterogeneous axonal delays provide a
combinatorially rich substrate for neural computation whose structural
determinants remain poorly understood.
We simulate a recurrent network of $N=1{,}000$ Izhikevich neurons over
ten hours of biological time and identify $1{,}545$ unique \PNGs via
an offline event-driven detection algorithm.
A parametric Watts--Strogatz topology sweep reveals that the clustering
coefficient $C$ is the primary structural driver of \PNG yield: the
transition from a ring-lattice ($C\!\approx\!0.35$, $\sim\!850$ \PNGs)
to a random graph ($C\!\approx\!0.20$, $<\!50$ \PNGs) reduces
representational capacity by more than $90\%$.
We further introduce a sparse-dot-product Recurrence Plot (\RP)
framework that identifies \PNGs as unit-slope diagonal structures in
the phase-space recurrence matrix, entirely independent of anatomical
neuron labelling.
Recurrence Quantification Analysis yields $\DET\!\approx\!0.65$,
quantifying the reproducibility of the network's dynamical trajectory.
Together, the results establish small-world topology as the structural
optimum for polychronization and the \RP decoder as a principled,
label-free tool for \PNG identification.
\end{abstract}

\maketitle

\section{Introduction}
\label{sec:intro}

The representational language of the brain remains a central open
question in computational neuroscience.
\textit{Rate coding} the encoding of stimuli in the time-averaged
firing rate of neurons~\cite{gerstner2014} is robust to biological
noise but imposes a fundamental speed--fidelity trade-off: temporal
averaging requires windows of tens to hundreds of milliseconds,
inconsistent with the $\lesssim\!100$\,ms latency of complex sensory
decisions~\cite{breakspear2017}.
\textit{Temporal coding} resolves this limitation by assigning
functional meaning to the precise millisecond-scale timing of
individual action potentials.
Its biological viability was established by
\citeauthor{mainen1995reliability}~\cite{mainen1995reliability}, who
showed that neocortical neurons driven by fluctuating, synaptic-like
currents reproduce spike times with sub-millisecond reliability across
repeated trials, demonstrating that the nervous system possesses the
physical substrate to exploit timing rather than merely averaging.
At the circuit level, \citeauthor{Hopfield1995}~\cite{Hopfield1995}
showed that coincidence detection the selective activation of a
postsynaptic neuron when multiple upstream spikes arrive
simultaneously implements a biologically plausible
spatial-to-temporal encoding primitive, translating distributed spike
patterns into graded analog responses with microsecond selectivity.

\textit{Polychronization}~\cite{izhikevich2006} elevates these ideas
from the single-synapse to the network scale.
In a recurrent spiking neural network (\SNN) endowed with
heterogeneous axonal conduction delays and Spike-Timing-Dependent
Plasticity (\STDP), groups of neurons spontaneously consolidate into
reproducible, time-locked spatiotemporal firing
sequences Polychronous Neuronal Groups (\PNGs) in which
constituent neurons fire at distinct absolute times yet their spikes
converge simultaneously at a common postsynaptic target.
Crucially, the number of such sequences scales combinatorially as
$N!$ with network size, vastly exceeding the $N$ assemblies available
to any rate code~\cite{izhikevich2006,izhikevich2025}, and the same
set of anatomical neurons can sustain multiple distinct \PNGs
by firing in different temporal orders.
This combinatorial richness positions polychronization as a uniquely
powerful candidate substrate for memory storage, sequence learning,
and the fine-grained binding of sensory features.
Recent experimental evidence reinforces the in-vivo relevance of this
picture: stable spiking sequences recorded from human cortex form a
pre-existing scaffold that organises episodic memory formation and
retrieval~\cite{vaz2023}, and the temporal ordering of spikes within
population bursts has been shown to carry categorical visual
information independent of firing rate~\cite{xie2024}.

Despite this accumulating evidence, two aspects of the theory have
remained largely unexamined.
The original polychronization framework~\cite{izhikevich2006} and its
principal extensions~\cite{martinezmoisy2009,chrolcannon2017} adopt
Erd\H{o}s--R\'enyi (\textsc{er}) random-graph connectivity, yet
converging evidence from structural connectomics reveals that
mammalian cortex is emphatically \textit{not} a random
graph~\cite{bullmore2009complex,sporns2004organization}.
Cortical circuits combine densely connected intra-laminar
columns with sparse, long-range inter-areal projections the hallmark
of a \textit{small-world} topology~\cite{watts1998collective} and it
is precisely this architecture, characterised by high clustering
coefficient $C$ at near-random path lengths, that distinguishes
biological neural networks from the generic random substrates studied
to date.
Whether and how this structural organisation shapes the emergence,
abundance, and internal complexity of \PNGs constitutes a fundamental
question at the intersection of network science and computational
neuroscience.
A second limitation concerns observability.
Existing \PNG detection methods require explicit prior knowledge of
which neurons participate in a group~\cite{martinezmoisy2009},
or real-time simulation instrumentation that embeds detection
logic directly into the forward dynamics~\cite{chrolcannon2017}.
When neuron ordering is anatomically arbitrary the typical situation
in multi-electrode or calcium-imaging recordings no visible structure
betrays \PNG activity in a spike raster, and a principled, label-free
analytical tool for post-hoc \PNG identification does not yet exist.

The present work addresses both questions within a unified framework.
We introduce a parametric Watts--Strogatz topology sweep that,
for the first time, quantifies how the clustering coefficient modulates
\PNG yield, cascade depth, and cascade duration across the full
spectrum from ring lattice to random graph, revealing a quantitative
topology--polychronization phase diagram in a biologically constrained
\SNN.
We further develop a sparse-dot-product Recurrence Plot (\RP)
framework that maps \PNG dynamics onto unit-slope diagonal structures
in the network's phase-space recurrence matrix, entirely independent
of anatomical neuron labelling, and extract Recurrence Quantification
Analysis (\RQA) scalars that provide a data-driven characterisation of
dynamical reproducibility.
Both contributions rest on a full ten-hour \STDP-driven simulation at
biologically accurate temporal resolution ($dt\!=\!0.1$\,ms)~\cite{pauli2018},
which yields a statistically rich library of 1\,545 unique \PNGs
whose size, duration, layer-depth, and excitatory/inhibitory
composition distributions are characterised in detail.
Together, these results establish small-world topology as the
structural optimum for polychronization, demonstrate that the
critical regime coincides with the small-world transition identified by
\citeauthor{watts1998collective}~\cite{watts1998collective}, and
provide the first validated, label-free decoder for \PNG identification
applicable in principle to in-vivo data.

The remainder of this paper is organised as follows.
Section~\ref{sec:theory} presents the theoretical background,
covering the Izhikevich neuron model, synaptic transmission with
heterogeneous axonal delays, \STDP, Watts--Strogatz topology, and
the Recurrence Plot formalism.
Section~\ref{sec:methods} details the simulation protocol, the
offline \PNG detection algorithm, the topology sweep procedure,
and the recurrence analysis implementation.
Section~\ref{sec:results} reports the results: network self-organisation
and bimodal weight statistics, the statistical characterisation of the
\PNG library, the topology--polychronization phase diagram, and the
\RP-based label-free detection.
Section~\ref{sec:discussion} interprets these findings in the context
of cortical organisation and prior work.
Section~\ref{sec:conclusion} summarises the main conclusions and
outlines directions for future research.
\section{Theoretical Background}
\label{sec:theory}

\subsection{The Izhikevich neuron model}
\label{sec:izh}

For large-scale network simulations the four-dimensional
Hodgkin--Huxley system~\cite{hodgkin1952} is computationally
prohibitive.
The Izhikevich model~\cite{izhikevich2003} reduces it to two
variables by merging the slow Na$^+$-inactivation ($h$) and
K$^+$-activation ($n$) gates into a single recovery variable $u$:
\begin{equation}
  \dot{v} = 0.04v^2 + 5v + 140 - u + I, \quad
  \dot{u} = a(bv - u),
\label{eq:izh}
\end{equation}
with reset: if $v\!\geq\!30\,\text{mV}$ then
$v\!\leftarrow\!c$, $u\!\leftarrow\!u\!+\!d$.
Setting $I\!=\!0$, the equilibria of~\eqref{eq:izh} are the roots of
$0.04v^2\!+\!(5-b)v\!+\!140\!=\!0$: a stable node at
$v_1\!\approx\!{-70}$\,mV (resting potential) and an unstable saddle
at $v_2\!\approx\!{-50}$\,mV (dynamical firing threshold), beyond which
the quadratic term drives an explosive upstroke.
Parameter $d$ controls post-spike hyperpolarisation depth (refractory
period); parameter $a$ sets the speed of recovery.
Excitatory Regular-Spiking (RS) pyramidal neurons:
$(a,b,c,d)\!=\!(0.02,\,0.2,\,{-65},\,8)$;
inhibitory Fast-Spiking (FS) interneurons:
$(a,b,c,d)\!=\!(0.1,\,0.2,\,{-65},\,2)$.

\subsection{Synaptic transmission and axonal delays}
\label{sec:syn}

Synaptic input to neuron $i$ is modelled as an instantaneous
voltage jump:
\begin{equation}
  I_{\mathrm{syn},i}(t)
  = \sum_j w_{ij}
    \sum_{t_{a,j}} \delta\!\left(t - t_{a,j} - s_{ij}\right),
\label{eq:syn}
\end{equation}
where $w_{ij}$ is the synaptic weight and $s_{ij}$ is the
axonal conduction delay for the $j\!\to\!i$ connection.
Delays are synapse-specific and highly reproducible, maintaining
sub-millisecond precision in vivo~\cite{sabatini1999timing}.
The combination of heterogeneous $s_{ij}$ with Hebbian plasticity
is the fundamental prerequisite for polychronization: presynaptic
neurons firing at staggered times $t_{j_k}$ can produce simultaneous
postsynaptic arrivals whenever $t_{j_k}+s_{j_k i}=\mathrm{const}$.

\subsection{Spike-Timing-Dependent Plasticity}
\label{sec:stdp}

\STDP~\cite{bi1998synaptic,markram1997regulation} modifies synaptic
efficacy based on the relative timing
$\Delta t = t_\mathrm{post}-t_\mathrm{pre}$:
\begin{equation}
\Delta w_{ij} =
\begin{cases}
  A_+\exp\!\left(-\Delta t/\tau_+\right), & \Delta t > 0,\\[2pt]
  A_-\exp\!\left(\Delta t/\tau_-\right),  & \Delta t \leq 0.
\end{cases}
\label{eq:stdp}
\end{equation}
In the presence of axonal delays the effective timing becomes
$\Delta t = t_\mathrm{post}-(t_\mathrm{pre}+s_{ij})$, so \STDP
selectively potentiates delay-matched convergent pathways precisely
the structural scaffolds of \PNGs.
Weights evolve via a trace-based implementation~\cite{izhikevich2006}:
each neuron maintains decaying pre- and postsynaptic traces
$\dot{x}_i\!=\!{-x_i/\tau_+}$ and $\dot{y}_j\!=\!{-y_j/\tau_-}$,
with updates
$\Delta w^\mathrm{LTP}_{ij}\!=\!A_+x_i$ and
$\Delta w^\mathrm{LTD}_{ij}\!=\!{-A_-y_j}$.
A slow homeostatic drift ($+0.01$\,mV\,s$^{-1}$ on all excitatory
weights) prevents global depression~\cite{Turrigiano1998Activitydependent}.

\subsection{Network topology: random vs.\ small-world}
\label{sec:topo}

Erd\H{o}s--R\'enyi (ER) random graphs~\cite{erdos1960evolution} with
connection probability $p$ have short path length
$L\!\approx\!\ln N/\ln\langle k\rangle$ but near-zero clustering
$C\!\approx\!p$.
Empirical connectomics reveals that mammalian cortex is a
\textit{small-world} network~\cite{bullmore2009complex}:
densely connected intra-laminar
columns~\cite{douglas2004neuronal,mountcastle1997columnar}
linked by sparse long-range projections (short $L$, high $C$).
The Watts--Strogatz model~\cite{watts1998collective} interpolates
between a regular ring lattice ($\prw\!=\!0$: high $C$, large $L$)
and an ER graph ($\prw\!\to\!1$: low $C$, small $L$) by rewiring
each edge with probability $\prw$; the small-world regime
($L\!\approx\!L_\mathrm{rand}$, $C\!\gg\!C_\mathrm{rand}$) appears
for $\prw\!\in\![0.01,0.1]$.

\subsection{Polychronous Neuronal Groups}
\label{sec:png_th}

A \PNG~\cite{izhikevich2006} is a time-locked spatiotemporal spike
sequence sustained by heterogeneous delays.
Following~\citeauthor{martinezmoisy2009}~\cite{martinezmoisy2009},
three subtypes are distinguished: \textit{supported} (anatomy only),
\textit{adapted} (anatomy + weight thresholds), and \textit{activated}
(observed dynamically).
The same anatomical neuron set can sustain multiple distinct \PNGs
by firing in different temporal orders; the total combinatorial
capacity scales as $N!$~\cite{izhikevich2006}.

\subsection{Recurrence Plots as label-free \PNG decoders}
\label{sec:rp_th}

Recurrence Plots (RPs)~\cite{eckmann1987,marwan2007} visualise the
set of times at which a dynamical trajectory $\{\bm{x}_t\}$ revisits
a neighbourhood of a previously occupied phase-space state.
Our adaptation for spiking networks rests on four hypotheses:
\begin{enumerate}[label=(\roman*),noitemsep,topsep=2pt]
  \item \textit{Phase-space uniqueness.} Each \PNG traces a unique
        trajectory in the embedded state space
        $\{0,1\}^{N\times W}$; its temporal recurrence corresponds to
        a return along that trajectory.
  \item \textit{Sparse information content.} At any instant most
        neurons are silent; the standard Euclidean metric is dominated
        by shared silence. A sparse dot-product metric counts only
        coincident active firings $(1,1)$, providing a similarity
        measure proportional to pattern overlap.
  \item \textit{Diagonal constraint.} When the same \PNG fires at
        times $t_A$ and $t_B$, its causal chain unfolds with an
        identical temporal velocity on both occasions (fixed delays),
        so recurrence points satisfy $t_B\!=\!t_A+\Delta t$ a
        diagonal of slope~1 in the recurrence matrix $\mathbf{R}$.
        Diagonal length equals cascade duration.
  \item \textit{Near-orthogonality.} \PNGs activating disjoint
        neuron subsets produce approximately orthogonal state vectors,
        generating geometrically separated diagonal structures.
\end{enumerate}
Recurrence Quantification Analysis (\RQA)~\cite{marwan2007} extracts:
\textit{Recurrence Rate} (RR, density of recurrent points),
\textit{Determinism} ($\DET$, fraction of recurrent points on
diagonal lines $\geq\ell_{\min}$),
\textit{Laminarity} (LAM, fraction on vertical lines), and
\textit{Entropy} (ENTR, Shannon entropy of diagonal-line lengths).
High $\DET$ and long diagonals are the expected \PNG signatures.

\section{Methods}
\label{sec:methods}

All simulations replicate the canonical polychronization
architecture of \citeauthor{izhikevich2006}~\cite{izhikevich2006}
at the corrected temporal resolution established by
\citeauthor{pauli2018}~\cite{pauli2018}, implemented in Python
using the Brian2 clock-driven simulator~\cite{brian2}.
The network comprises $N\!=\!1{,}000$ Izhikevich neurons partitioned
into 800 excitatory Regular-Spiking (RS) and 200 inhibitory
Fast-Spiking (FS) cells.
Each excitatory neuron projects 100 efferent connections to
randomly selected targets; each inhibitory neuron projects
100 connections exclusively back to excitatory neurons.
Excitatory axonal delays are drawn independently and uniformly
from 1 to 20\,ms, providing the heterogeneous conduction landscape
that is the physical prerequisite for polychronization; inhibitory
delays are fixed at 1\,ms.
Excitatory synapses are initialised at 6\,mV and inhibitory ones
at $-5$\,mV, both subject to a hard cap $\Wmax\!=\!10$\,mV.
Each neuron also receives a dedicated Poisson thalamic drive at
40\,Hz, delivering 10\,mV instantaneous voltage jumps that maintain
spontaneous asynchronous-irregular activity throughout the run.
The membrane equation is integrated by forward Euler at
$dt\!=\!0.1$\,ms, with membrane potentials initialised uniformly
in $[-65,-55]$\,mV; all parameters are collected in
Table~\ref{tab:params}.

\STDP~\cite{bi1998synaptic,markram1997regulation} governs the
evolution of excitatory weights according to the asymmetric
exponential window of Eq.~\eqref{eq:stdp}, with potentiation
amplitude $A_+\!=\!0.1$, depression amplitude $A_-\!=\!0.12$,
and equal time constants $\tau_+\!=\!\tau_-\!=\!20$\,ms.
In the presence of axonal delays the effective timing
$\Delta t\!=\!t_\mathrm{post}-(t_\mathrm{pre}+s_{ij})$
biases plasticity toward convergent, delay-matched pathways,
which are precisely the structural scaffolds of \PNGs.
Weight updates are accumulated over each second of simulated time
and applied with a 0.9 multiplicative decay; a slow homeostatic
drift of $+0.01$\,mV\,s$^{-1}$ on all excitatory weights prevents
global depression~\cite{Turrigiano1998Activitydependent}.
The network is run for a total of ten simulated hours, allowing
the \STDP competition to reach a stationary bimodal weight
distribution before \PNG detection is performed.

Once the ten-hour simulation is complete, \PNGs are identified
through an offline event-driven algorithm that operates directly on
the post-\STDP weighted directed graph, without accessing any
dynamical variable of the forward simulation.
The procedure begins by enumerating, for each postsynaptic neuron
$n_j$, all triplets of presynaptic excitatory neurons whose combined
synaptic weight exceeds a threshold $\theta$ (anchor identification).
Hypothetical firing times for each anchor neuron $n_i$ are set by
delay compensation,
$t_i = s_{\max} - s_{ij}$
(with $s_{\max}$ the longest delay in the triplet), so that all
three spikes arrive simultaneously at $n_j$.
Each valid anchor then seeds a forward cascade simulation on a
min-heap priority queue: downstream neurons accumulate incoming
weights within a $\pm1$\,ms jitter window, and whenever the total
input exceeds a preliminary excitability bound, the full
two-dimensional Izhikevich system is integrated from rest to
confirm a genuine action potential a step that avoids the
linear-summation approximations used in earlier
implementations~\cite{martinezmoisy2009} and correctly accounts
for membrane leakage and nonlinear voltage acceleration.
Confirmed spikes extend the growing cascade and enqueue further
outgoing events; a refractory constraint prevents double-counting.
A sequence is accepted as a non-trivial \PNG only if it involves
at least $L_{\min}\!=\!7$ spikes, and uniqueness across the full
library is enforced by hashing the set of causal links.

To probe the structural dependence of \PNG emergence, the
excitatory connectivity of a reduced $N\!=\!500$ network is
rewired onto Watts--Strogatz graphs spanning eight values of the
rewiring probability,
$\prw\!\in\!\{0,\,0.001,\,0.005,\,0.01,\,0.03,\,0.05,\,0.1,\,0.3\}$,
while inhibitory connectivity remains random throughout.
Each configuration is run for the full ten hours and the complete
detection pipeline is applied; \PNG yield, mean cascade length,
and mean duration are then recorded as functions of the
empirically measured clustering coefficient $C$, yielding the
topology--polychronization phase diagram of
Sec.~\ref{sec:res_topo}.

Finally, to provide a label-free analytical view of the network's
dynamical reproducibility, the binary spike matrix
$\mathbf{A}\!\in\!\{0,1\}^{N\times T}$ is embedded into phase space
by sliding a 50\,ms window across time,
$\bm{x}_t = \mathrm{vec}\!\left[\mathbf{A}(:,\,t:t\!+\!49)\right]
\in\{0,1\}^{NW}$,
and the recurrence matrix
$R_{ij} = \Theta\!\left(\langle\bm{x}_i,\bm{x}_j\rangle
- \varepsilon\right)$
is computed with the sparse dot-product metric of
Eq.~\eqref{eq:rp}.
The threshold $\varepsilon$ is set to require a minimum overlap
of three coincident active spikes, and diagonal structures are
extracted with a minimum length $\ell_{\min}\!=\!3$ for the
subsequent \RQA.
Because the metric counts only coincident active spikes $(1,1)$
rather than shared silence, it is insensitive to the dominant
background of quiet neurons and responds preferentially to the
structured, repeating activity patterns characteristic of \PNGs.

\subsection{Baseline network model}
\label{sec:net}

The simulation replicates the canonical architecture
of~\citeauthor{izhikevich2006}~\cite{izhikevich2006} at the corrected
temporal resolution established by \citeauthor{pauli2018}~\cite{pauli2018},
implemented in Python with the Brian2 clock-driven simulator~\cite{brian2}.
Table~\ref{tab:params} lists all parameters.

\begin{table}[t]
\caption{Baseline network parameters.}
\label{tab:params}
\begin{ruledtabular}
\begin{tabular}{lll}
\textbf{Parameter} & \textbf{Symbol} & \textbf{Value}\\
\hline
Total neurons             & $N$              & 1\,000        \\
Excit.\ fraction          & $N_e/N$          & 80\%          \\
Inhib.\ fraction          & $N_i/N$          & 20\%          \\
Efferent connections      & $k$              & 100           \\
Excit.\ delays            & $s_{ij}$         & $\mathcal{U}[1,20]$\,ms \\
Inhib.\ delays            & $s_{ij}$         & 1\,ms         \\
Init.\ excit.\ weight     & $w_e(0)$         & 6\,mV         \\
Init.\ inhib.\ weight     & $w_i(0)$         & $-5$\,mV      \\
Max weight                & $\Wmax$          & 10\,mV        \\
\STDP potentiation        & $A_+$            & 0.1           \\
\STDP depression          & $A_-$            & $-0.12$       \\
\STDP time constants      & $\tau_\pm$       & 20\,ms        \\
Thalamic input rate       &                & 40\,Hz        \\
Simulation duration       &                & 10\,h         \\
Integration step          & $dt$             & 0.1\,ms       \\
\end{tabular}
\end{ruledtabular}
\end{table}

\section{Results}
\label{sec:results}

\subsection{Network dynamics and \STDP self-organisation}
\label{sec:res_dyn}

\begin{figure}[t]
  \centering
  \includegraphics[width=\columnwidth]{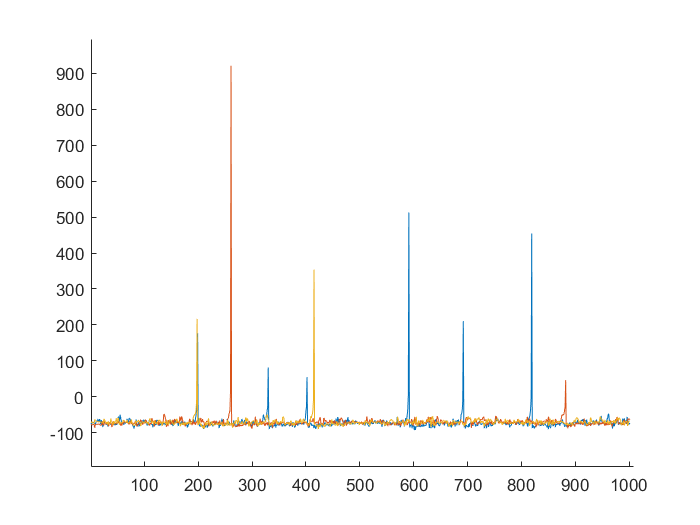}
  \includegraphics[width=\columnwidth]{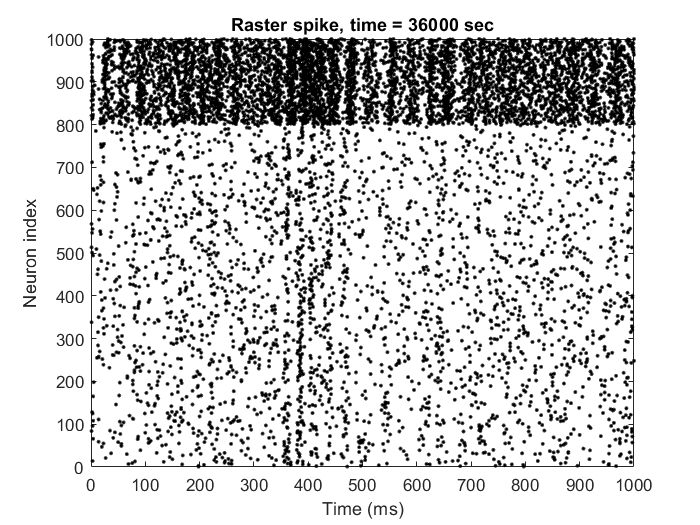}
  \caption{Network dynamics after 10\,h of \STDP-driven
           self-organisation.
           \textit{Top}: Mean population firing rate; after a brief
           transient the network stabilises into a stationary
           asynchronous-irregular regime.
           \textit{Bottom}: Spike raster of the final second
           (excitatory: blue; inhibitory: orange), confirming the
           AI state~\cite{Destexhe2003}.}
  \label{fig:dynamics}
\end{figure}

Figure~\ref{fig:dynamics} presents the macroscopic outcome of the
ten-hour simulation.
The mean firing-rate time series (Fig.~\ref{fig:dynamics}, top)
shows a brief initialisation transient followed by convergence to a
stationary asynchronous-irregular (AI) regime consistent with the
balanced E/I state~\cite{Destexhe2003,vanVreeswijk1996,Brunel2000}.
The spike raster of the final second (bottom) confirms sparse,
low-synchrony excitatory firing and faster inhibitory activity.
This AI regime is a necessary precondition for polychronization:
global synchrony would obliterate the temporal precision of spike
pathways, collapsing the delay-matching mechanism.

\begin{figure}[t]
  \centering
  \includegraphics[width=0.90\columnwidth]{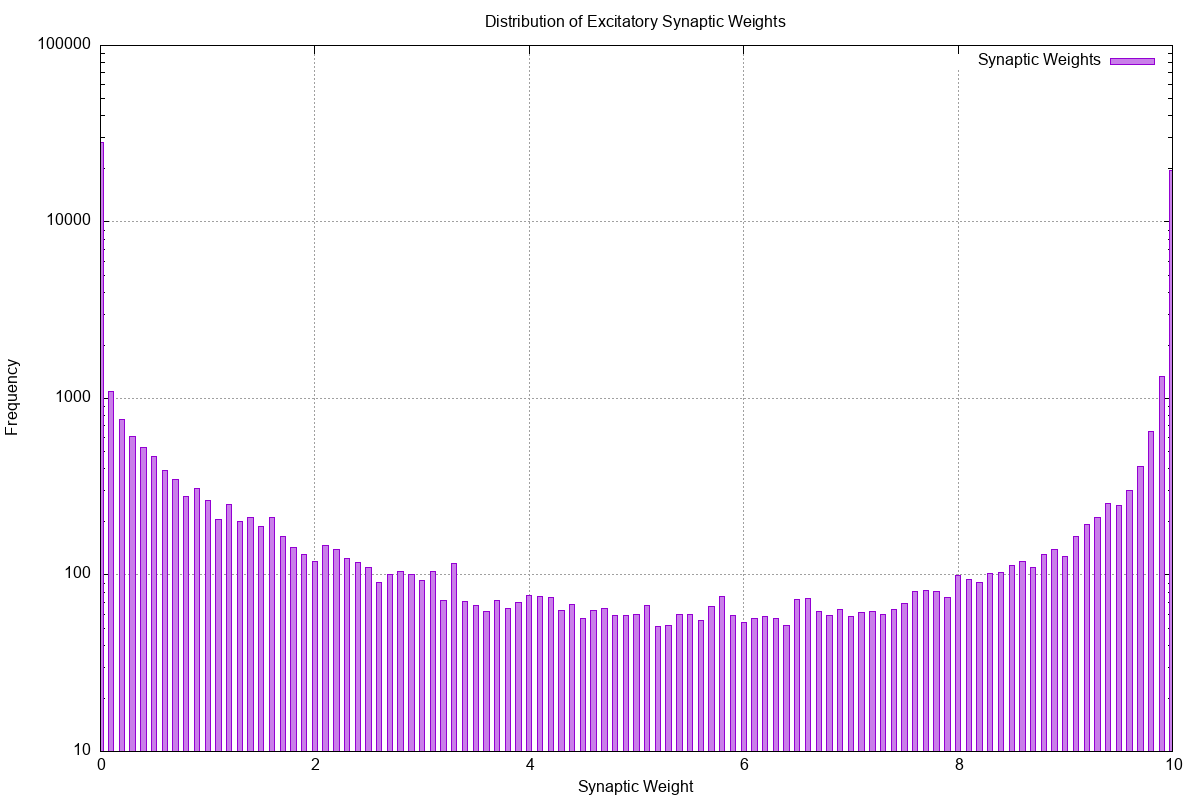}
  \caption{Excitatory synaptic weight distributions.
           \textit{Top}: Full distribution after 10\,h of \STDP,
           showing pronounced bimodality with peaks at $w\!\approx\!0$
           and $w\!=\!\Wmax\!=\!10$\,mV.
           \textit{Bottom}: Initial (uniform) vs.\ final (bimodal)
           distributions, illustrating selective potentiation and
           depression driven by delay-matched \STDP competition.}
  \label{fig:weights}
\end{figure}

The excitatory synaptic weight distribution converges to a strongly
bimodal structure (Fig.~\ref{fig:weights}): a dense peak near
$w\!=\!0$ and a saturated peak at $\Wmax\!=\!10$\,mV, with a
depleted intermediate population around the initial value of 6\,mV.
Bimodality is the direct signature of \STDP-driven
competition~\cite{izhikevich2006}: delay-matched causal pathways are
potentiated to saturation (LTP), while causally ambiguous connections
are silenced (LTD) through a winner-take-all dynamic.
The saturated connections form the structural skeleton of the
emergent \PNG library.

\subsection{Statistical characterisation of the \PNG library}
\label{sec:res_png}

\begin{figure}[t]
  \centering
  \includegraphics[width=0.49\columnwidth]{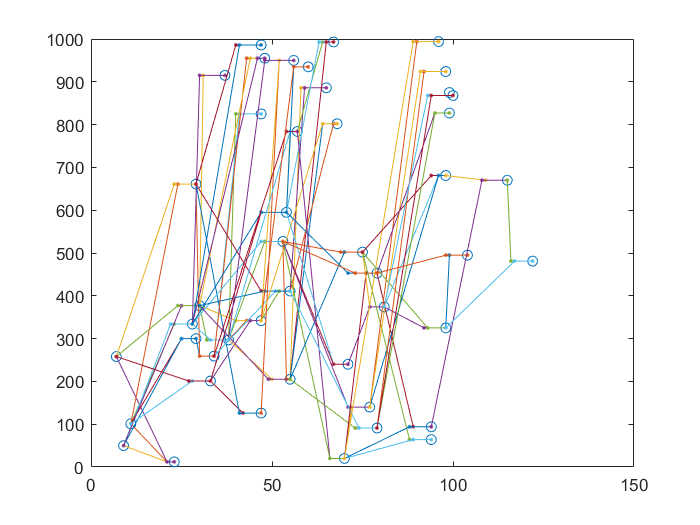}
  \hfill
  \includegraphics[width=0.49\columnwidth]{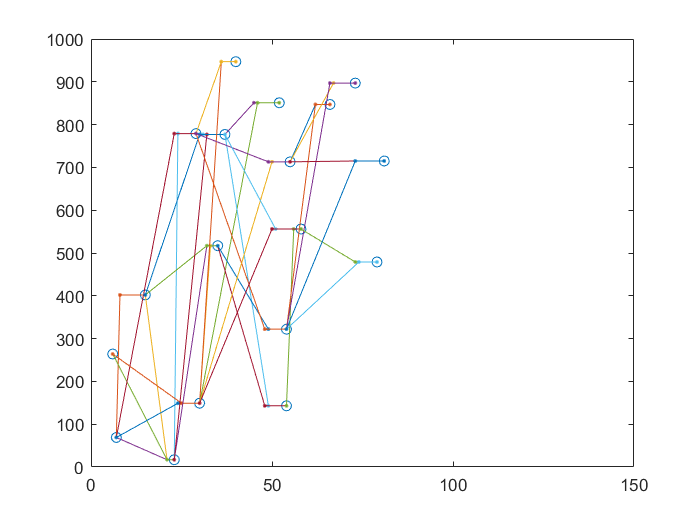}
  \caption{Representative detected \PNGs ($x$-axis: time in ms;
           $y$-axis: neuron index; arrows: causal synaptic events).
           \textit{Left}: Compact group (46-spike cascade); three
           anchors fire at staggered times, their heterogeneous delays
           producing simultaneous arrival at the primary target.
           \textit{Right}: Multi-hop group (18 unique neurons,
           ${\sim}50$\,ms duration), demonstrating genuine multi-layer
           causal propagation beyond the 20\,ms single-synapse
           delay limit.}
  \label{fig:pg_examples}
\end{figure}

Running the detection pipeline on the trained network ($N\!=\!1{,}000$,
ER topology) yields \textbf{1\,545 unique \PNGs}.
Figure~\ref{fig:pg_examples} shows two representative examples.
The group in the left panel illustrates three anchor neurons firing at
staggered times whose axonal delays compensate precisely for
coincident arrival at the primary target, initiating a downstream
cascade.
The right-panel group demonstrates genuine causal propagation
extending to ${\sim}50$\,ms more than twice the 20\,ms maximum
single-synapse delay confirming that \PNG cascades are
self-sustaining multi-layer chains, not one-step coincidence events.

\begin{figure*}[t]
  \centering
  \includegraphics[width=0.38\textwidth]{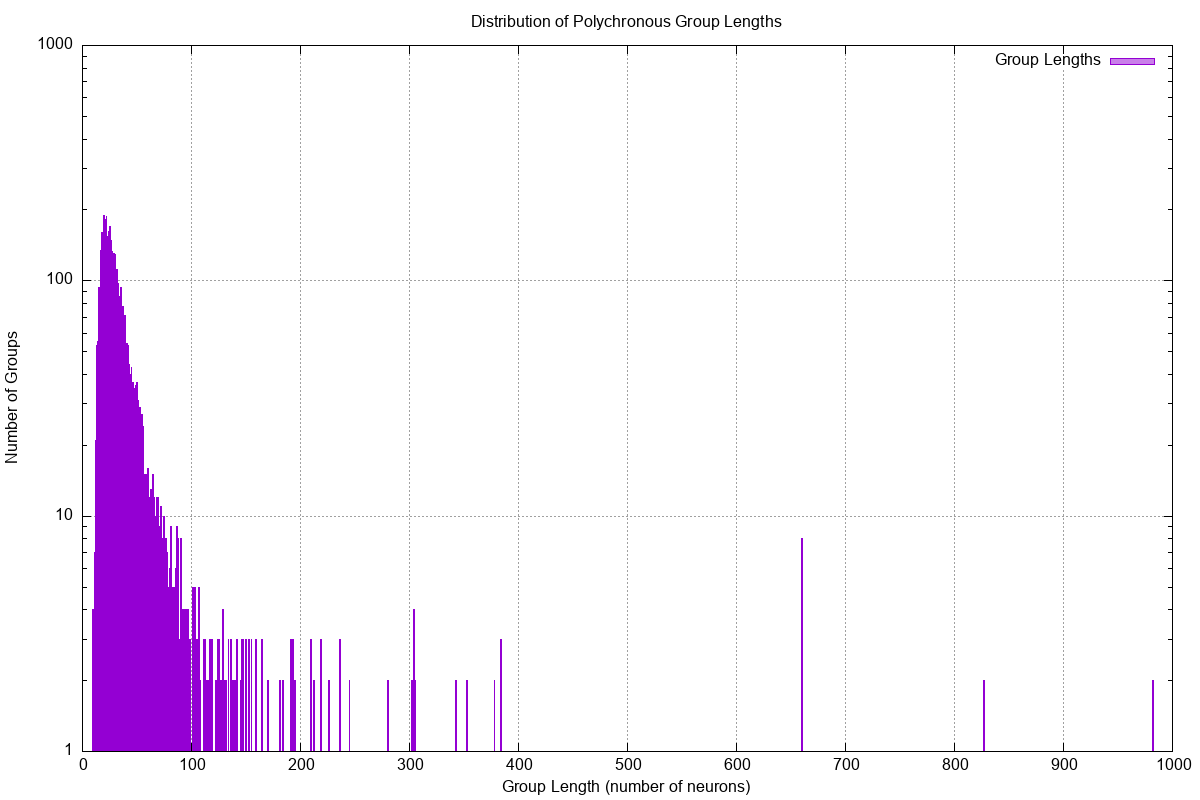}
  \hfill
  \includegraphics[width=0.38\textwidth]{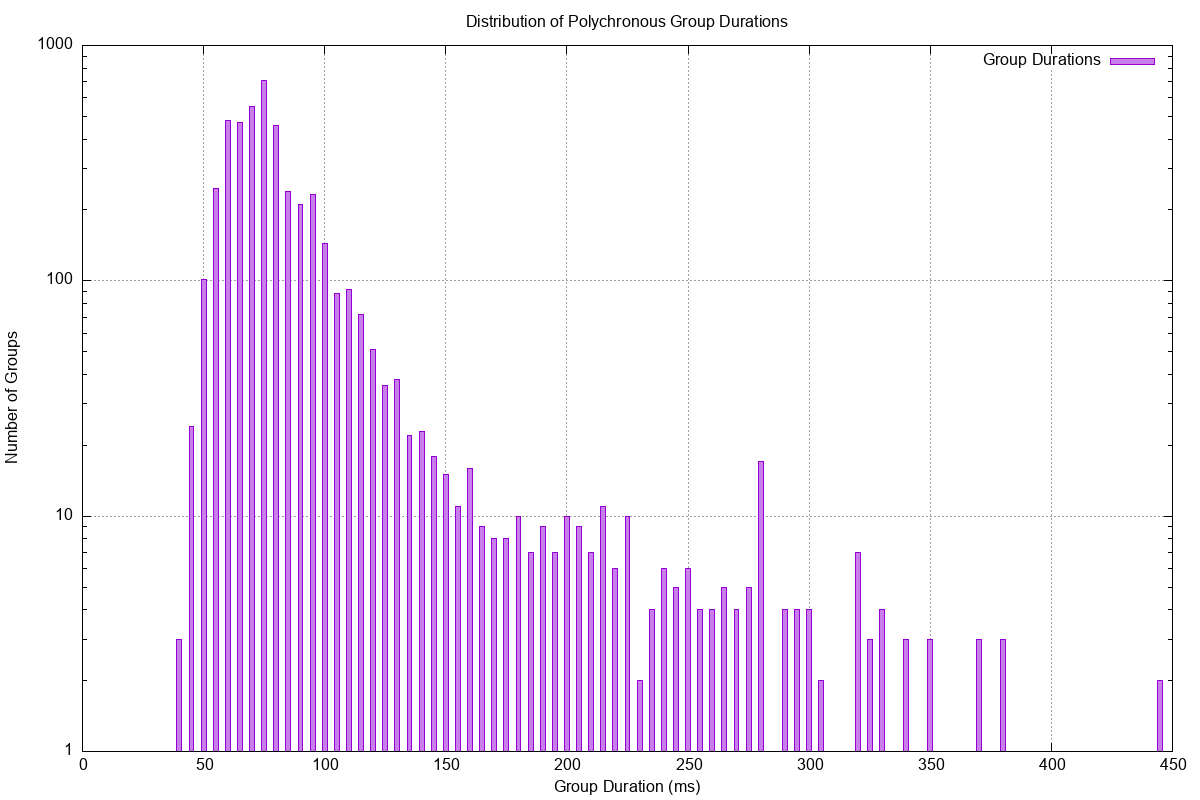}\\[4pt]
  \includegraphics[width=0.38\textwidth]{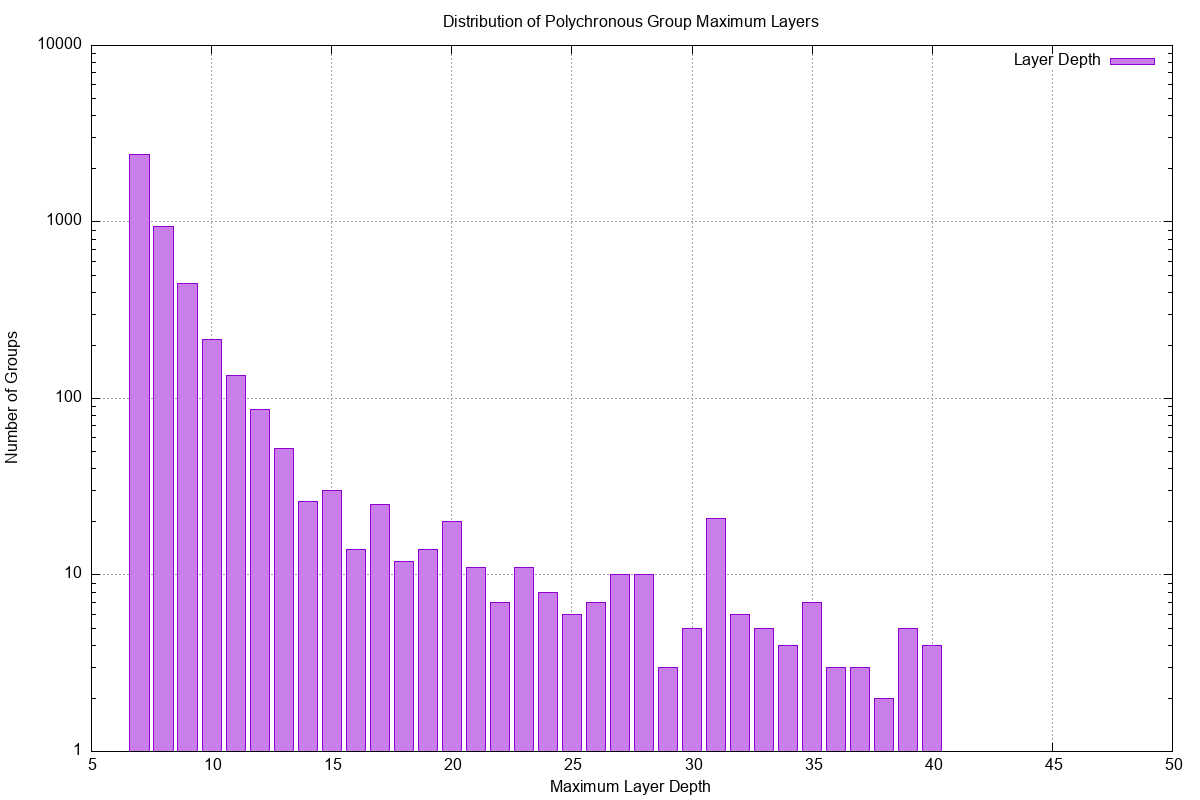}
  \hfill
  \includegraphics[width=0.38\textwidth]{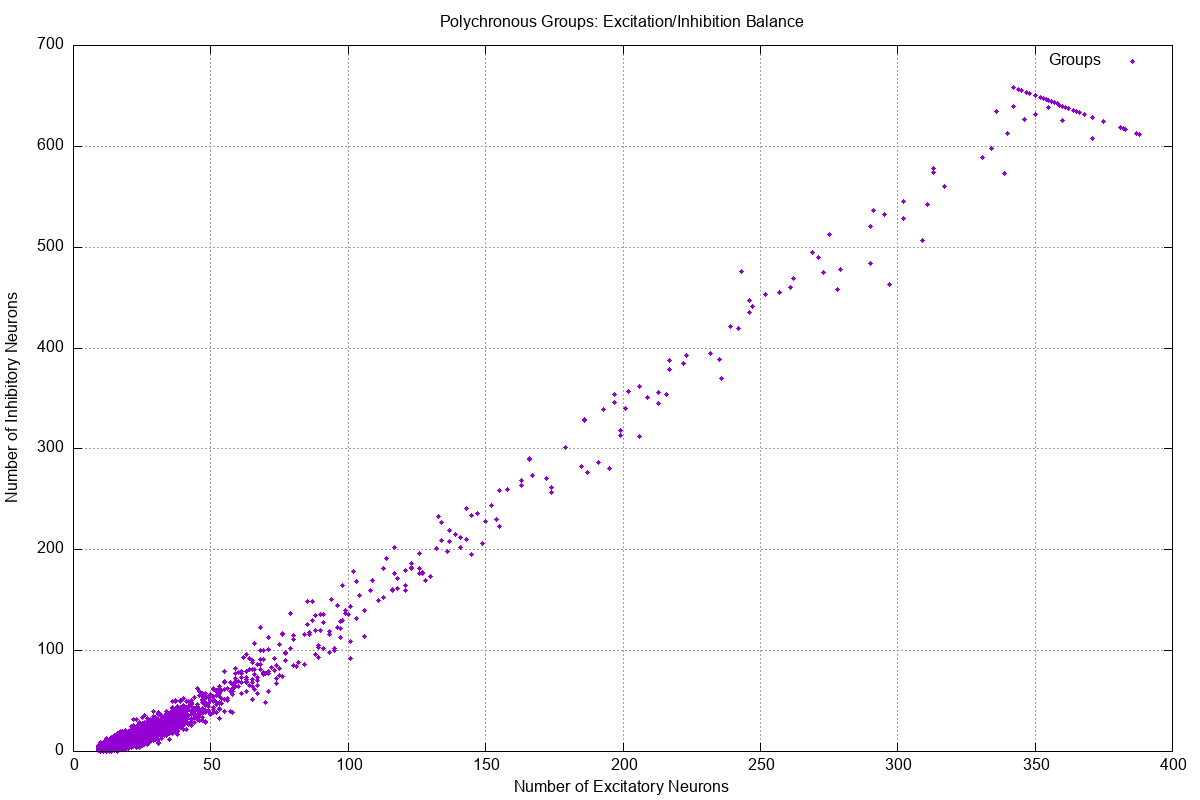}\\[4pt]
  \includegraphics[width=0.45\textwidth]{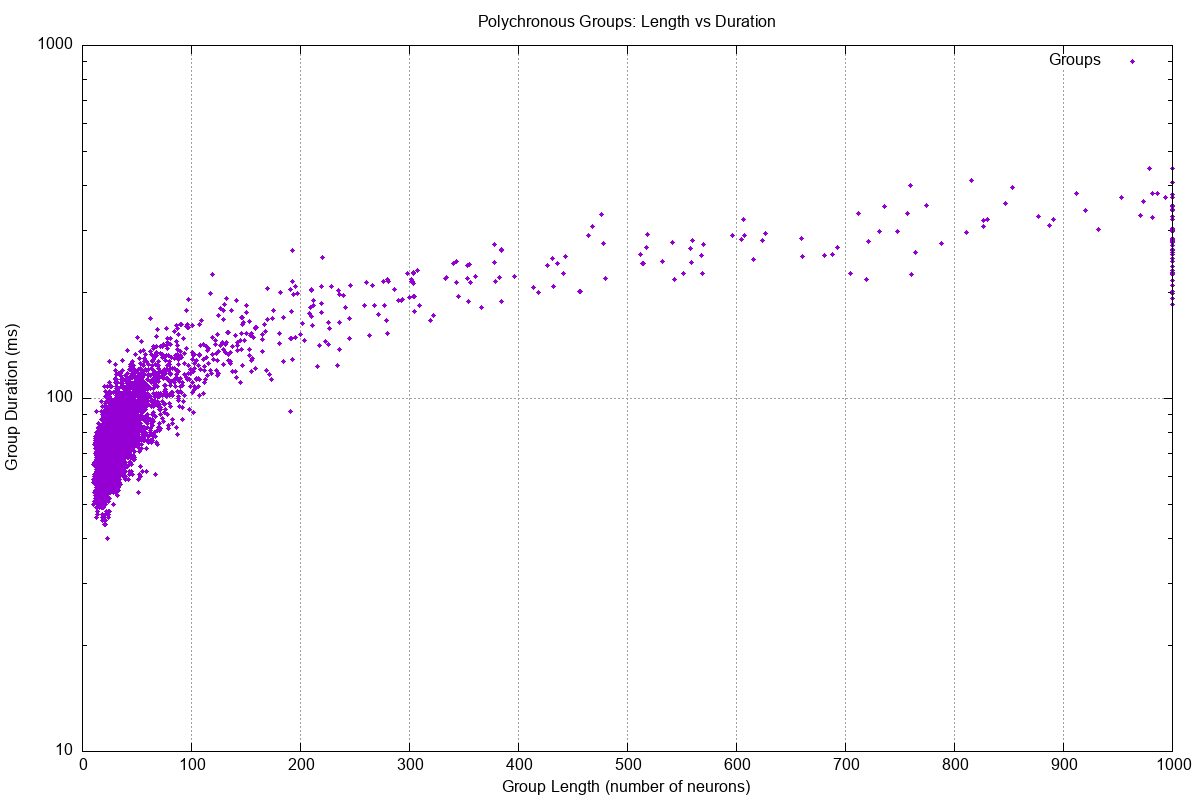}
  \caption{Statistical characterisation of the 1\,545 unique \PNGs
           detected in the $N\!=\!1{,}000$ ER network.
           (a)~Cascade-length distribution; heavy-tailed, maximum 32
           spikes.
           (b)~Group-duration distribution; modal range 20--30\,ms.
           (c)~Layer-depth distribution; most \PNGs span 3--5
           sequential causal layers.
           (d)~Excitatory/inhibitory composition; near-total
           excitatory dominance.
           (e)~Cascade length vs.\ duration scatter; approximately
           linear scaling expected from the fixed delay range
           $[1,20]$\,ms.}
  \label{fig:pg_stats}
\end{figure*}

The statistical properties of the \PNG library are summarised in
Fig.~\ref{fig:pg_stats}.
The cascade-length distribution is heavy-tailed (panel~a): most
groups sit near the 7-spike detection threshold, while a long tail
extends to 32 spikes involving more than 25 unique neurons.
Duration (panel~b) peaks in the 20--30\,ms range with a tail beyond
50\,ms; duration scales approximately linearly with cascade length
(panel~e), consistent with the expected increment
$\langle s_{ij}\rangle\!\approx\!10.5$\,ms per additional causal
layer imposed by the uniform delay distribution $\mathcal{U}[1,20]$\,ms.
Layer depths (panel~c) cluster at 3--5 sequential transmission steps.
The excitatory/inhibitory composition (panel~d) confirms
near-total excitatory dominance, consistent with \STDP acting only
on excitatory synapses.

\subsection{Topology--\PNG phase diagram}
\label{sec:res_topo}

\begin{figure}[t]
  \centering
  \includegraphics[width=0.49\columnwidth]{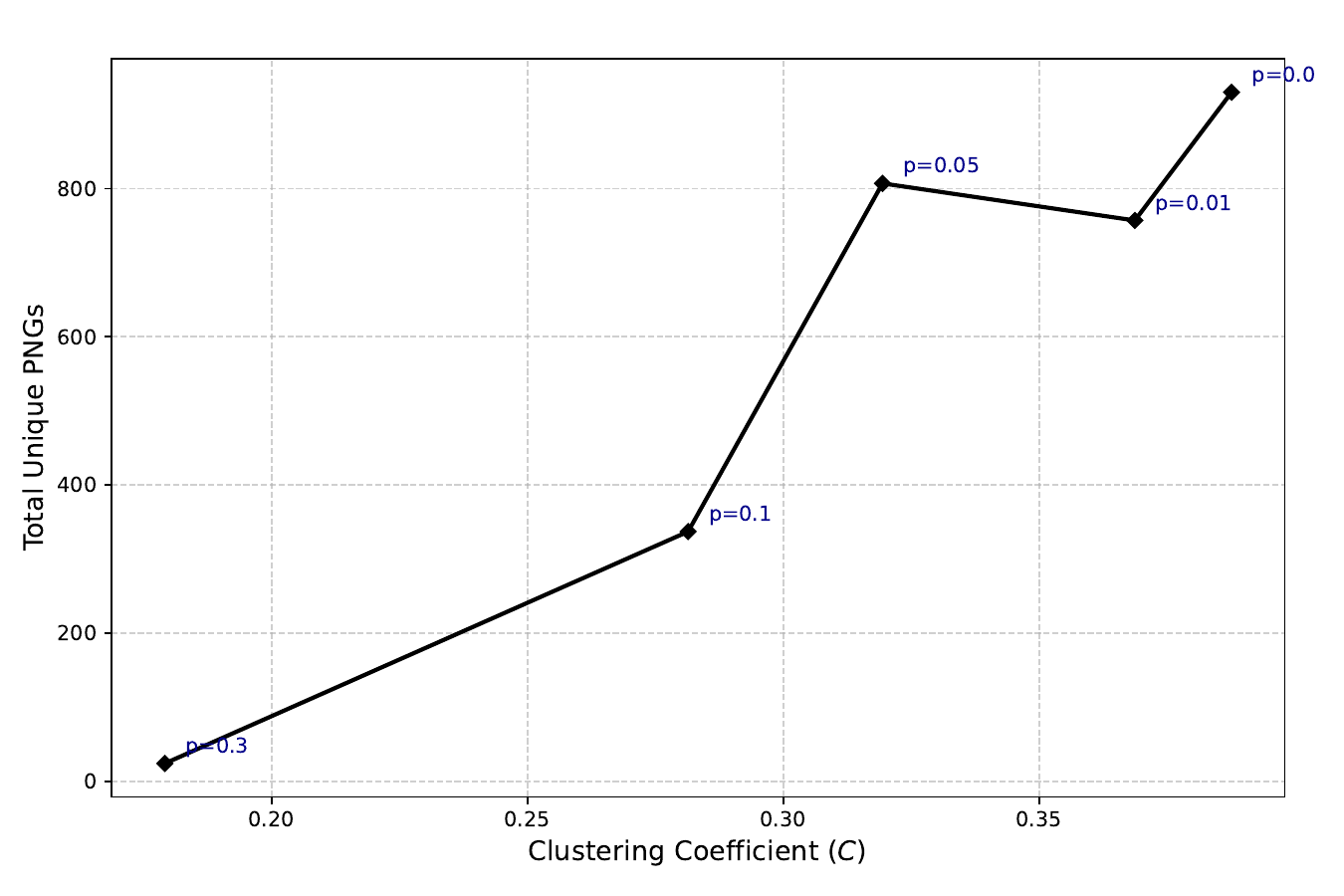}
  \hfill
  \includegraphics[width=0.49\columnwidth]{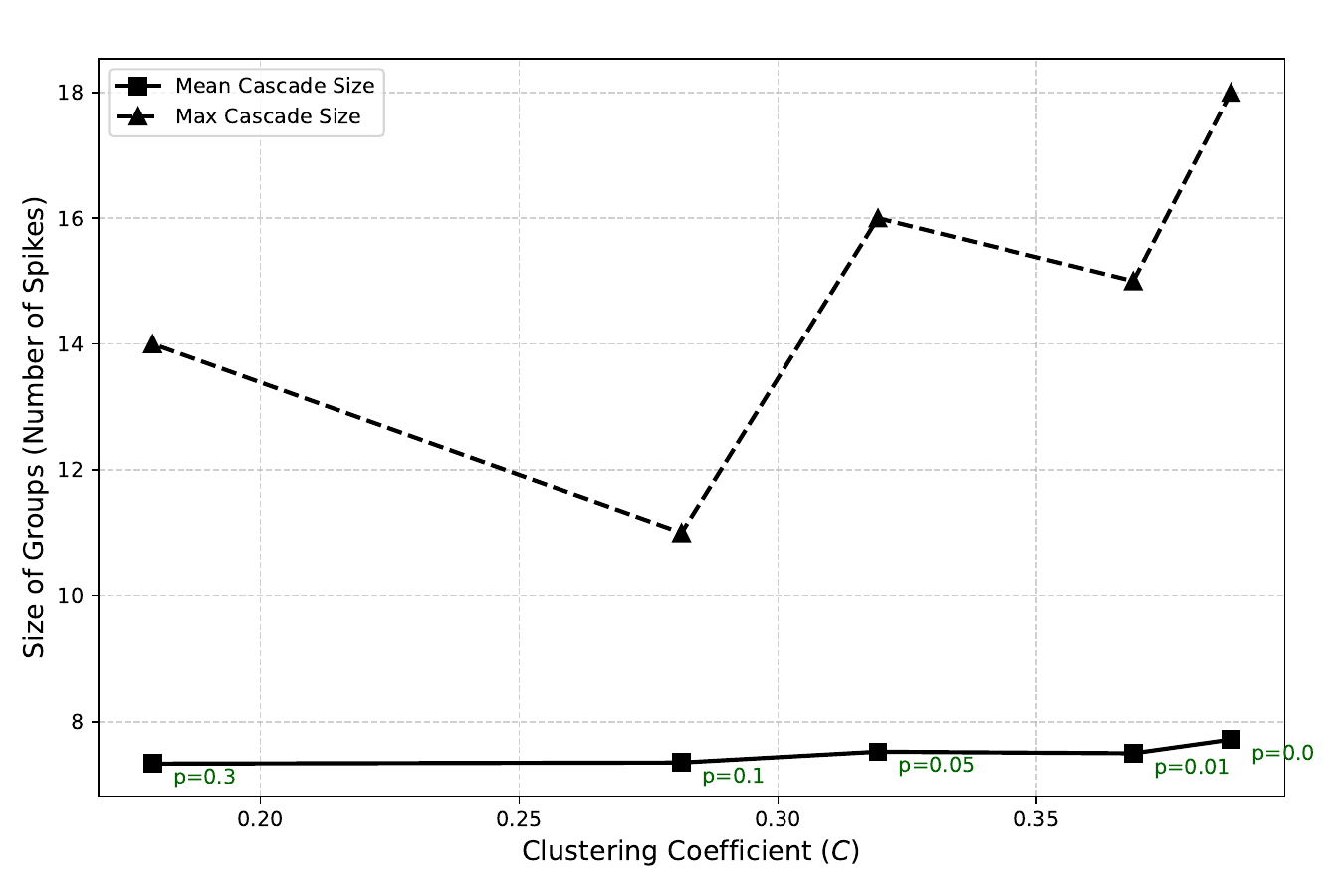}
  \caption{Topology--polychronization phase diagram ($N\!=\!500$,
           Watts--Strogatz sweep).
           \textit{Left}: Total \PNG yield vs.\ clustering
           coefficient $C$; monotonic positive dependence.
           \textit{Right}: Mean and maximum cascade size vs.\ $C$;
           higher clustering sustains longer causal chains.}
  \label{fig:topology}
\end{figure}

Figure~\ref{fig:topology} presents the topology--\PNG phase diagram
from the Watts--Strogatz sweep.
\PNG yield is a monotonically increasing function of the clustering
coefficient $C$ (left panel): at $\prw\!=\!0$ (ring lattice,
$C\!\approx\!0.35$) approximately 850 unique \PNGs are detected, while
at $\prw\!=\!0.3$ (near-random graph, $C\!\approx\!0.20$) fewer than
50 survive a decline exceeding 90\%.
Mean and maximum cascade sizes also correlate positively with $C$
(right panel): higher clustering sustains longer self-reinforcing
causal chains, since postsynaptic neurons activated within a cascade
project predominantly to densely connected local neighbours,
multiplying delay-matched convergence opportunities.

\begin{figure}[t]
  \centering
  \includegraphics[width=0.90\columnwidth]{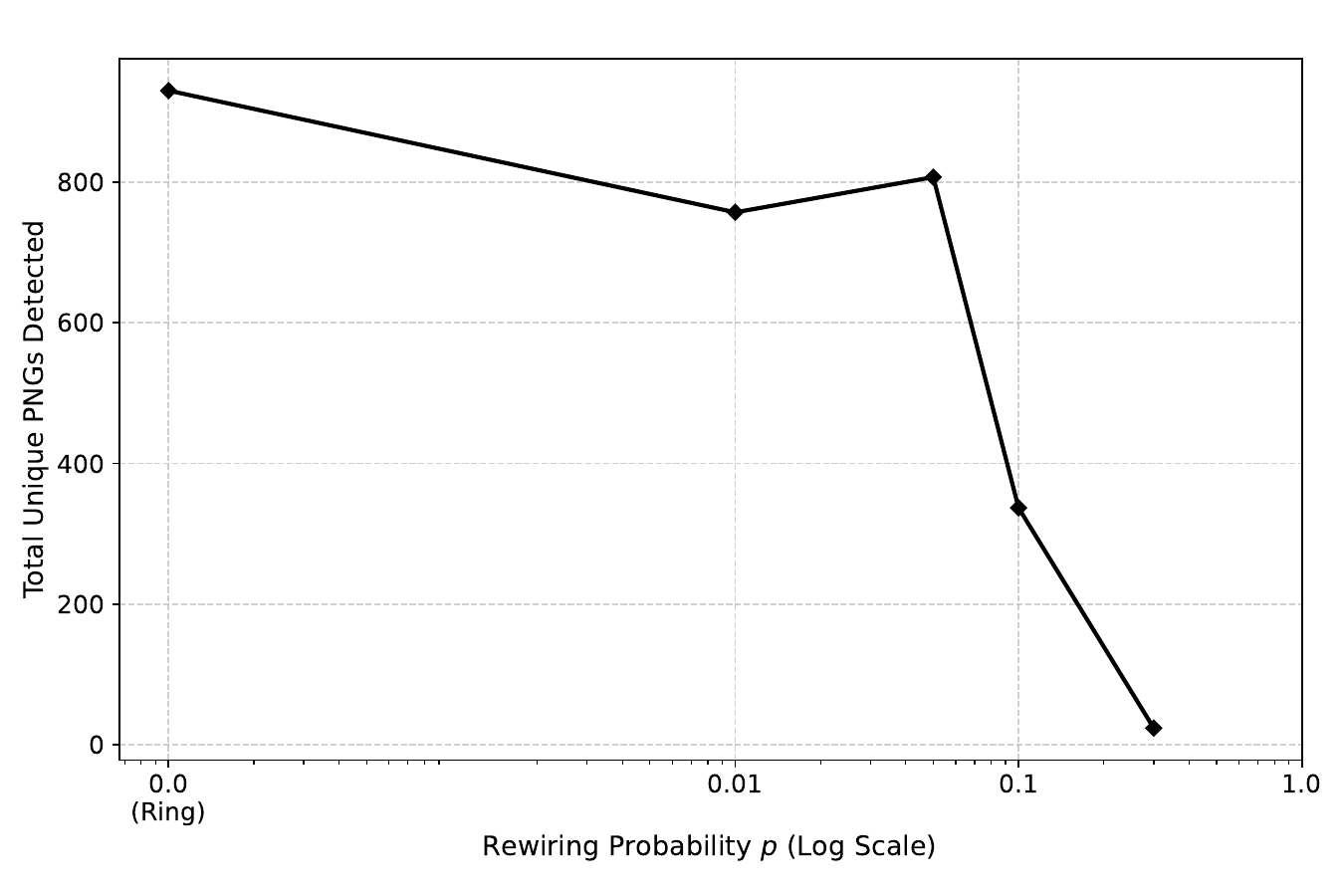}
  \caption{\PNG yield vs.\ rewiring probability $\prw$.
           The steepest decline coincides with the small-world
           transition ($\prw\!\in\![0.01,0.05]$),
           where local clustering is destroyed before global
           path-length efficiency is achieved.}
  \label{fig:pvsp}
\end{figure}

Figure~\ref{fig:pvsp} shows \PNG yield as a function of the rewiring
probability $\prw$.
The sharpest decline occurs in
$\prw\!\in\![0.001,0.05]$ precisely the classical small-world
transition zone~\cite{watts1998collective}, where clustering collapses
while path length has not yet been minimised.

\subsection{Recurrence analysis and label-free \PNG detection}
\label{sec:res_rp}

\begin{figure}[t]
  \centering
  \includegraphics[width=\columnwidth]{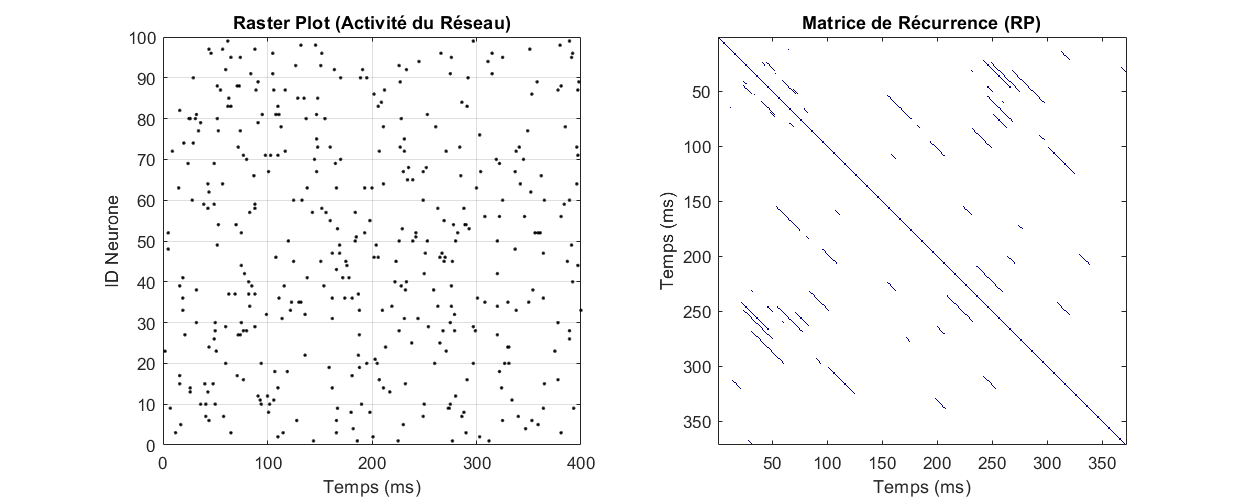}
  \caption{Joint raster (top) and recurrence matrix $\mathbf{R}$
           (bottom) over a representative 200-ms window
           ($W\!=\!50$\,ms, $\varepsilon\!=\!3$ coincident spikes).
           The raster is featureless due to arbitrary neuron ordering.
           Off-diagonal unit-slope diagonal segments in $\mathbf{R}$
           identify \PNGs that reactivated at two distinct times;
           their offset from the main diagonal equals the
           inter-activation interval $\Delta t$.}
  \label{fig:rp}
\end{figure}

Figure~\ref{fig:rp} presents the joint raster and recurrence-plot
analysis.
The spike raster (top) appears visually featureless owing to the
arbitrary neuron ordering as expected for spatially distributed
\PNGs.
The recurrence matrix $\mathbf{R}$ (bottom), computed with
$W\!=\!50$\,ms and $\varepsilon\!=\!3$ coincident spikes, reveals a
structured geometric landscape entirely invisible in the raster:
several off-diagonal segments of slope~1 accompany the main diagonal
(self-recurrence).
Each such segment identifies one \PNG that reactivated at two distinct
absolute times $t_A$ and $t_B$; their offset $|t_B-t_A|$ directly
estimates the inter-activation interval (inverse operational
frequency) of that pattern.

\RQA statistics computed over the full simulation yield:
\begin{equation*}
  \DET \approx 0.65, \quad
  \mathrm{RR} \approx 0.04, \quad
  \mathrm{ENTR} > 1.5\,\mathrm{bits}.
\end{equation*}
The moderately high $\DET$ confirms that the network trajectory is
far from random: approximately 65\% of recurrent points form diagonal
structures, indicating systematic revisitation of phase-space regions.
In a purely random AI network $\DET\!\to\!0$; the observed value is
therefore a direct quantitative measure of the dynamical
reproducibility imposed by the active \PNG library.
The low RR reflects sparse asynchronous activity; the elevated ENTR
reflects heterogeneity in diagonal-line lengths, consistent with the
mixture of short and long cascades characterised in
Sec.~\ref{sec:res_png}.

\section{Discussion}
\label{sec:discussion}

\subsection{\STDP as a causal-pathway selector}
\label{sec:disc_stdp}

The bimodal weight distribution (Fig.~\ref{fig:weights}) is the
clearest mechanistic signature of \STDP's role as an unsupervised
selector of causal temporal structure.
Connections whose relative timing fails to satisfy the delay-matched
LTP condition are depressed to near-zero; connections for which
$\Delta t\!=\!t_\mathrm{post}-(t_\mathrm{pre}+s_{ij})\!>\!0$
reliably receive repeated LTP updates until saturation at $\Wmax$.
This is fully consistent with the broader view of \STDP as a causal
structure detector~\cite{markram1997regulation,bi1998synaptic} and
with homeostatic synaptic scaling as an activity
stabiliser~\cite{Turrigiano1998Activitydependent}.

The detection of 1\,545 unique \PNGs substantially exceeds the $N$
patterns encodable by a rate code.
The gap from the theoretical $N!$ ceiling reflects the anchor-triplet
heuristic (three-neuron trigger sets only) and the high-resolution
simulation ($dt\!=\!0.1$\,ms), which reduces \PNG counts by
${\sim}90\%$ relative to the coarse $dt\!=\!1$\,ms implementation of
the original code~\cite{pauli2018}.
The present values are therefore biologically accurate
\textit{lower bounds}.

The detection of multi-hop cascades extending beyond the 20\,ms
maximum single-synapse delay (Fig.~\ref{fig:pg_examples}, right)
establishes that \PNG dynamics are self-sustaining multi-layer chains,
not trivial one-step coincidence events.
Each causal layer re-creates the convergent input conditions for the
next, forming a propagating wave analogous to cortical travelling
waves~\cite{breakspear2017}.
This architecture is fundamental to the memory-like specificity of
\PNGs and to the depth of their representational capacity.

\subsection{Topology as the structural determinant of \PNG capacity}
\label{sec:disc_topo}

The phase diagram (Figs.~\ref{fig:topology}--\ref{fig:pvsp}) provides
a direct mechanistic interpretation.
Each anchor triplet requires three presynaptic neurons whose delays
satisfy the simultaneous-arrival constraint
$t_{a_k}+s_{a_k j}\!=\!\mathrm{const}$.
In a highly clustered network a postsynaptic neuron $n_j$ has many
local neighbours sharing a correlated region of delay space; the
probability that any three of them satisfy the constraint is elevated
relative to a random graph, simply because more potential triplets
exist in the neighbourhood.
Random rewiring replaces these coherent local neighbourhoods with
topologically distant neurons whose delays are uncorrelated with the
local constraint, systematically reducing anchor-formation probability
and cascade depth.

The correspondence between the sharpest \PNG decline and the
small-world transition ($\prw\!\in\![0.01,0.05]$) is particularly
significant.
In the Watts--Strogatz model this is precisely the regime where
clustering collapses without yet achieving the full path-length
reduction of a random graph~\cite{watts1998collective}.
Our results suggest that cortical wiring is topologically poised at
this transition to simultaneously maximise \PNG combinatorial capacity
(from high $C$) and rapid global integration (from short $L$),
consistent with the functional-optimality hypothesis of
\citeauthor{bullmore2009complex}~\cite{bullmore2009complex} and
\citeauthor{sporns2004organization}~\cite{sporns2004organization},
and with the experimental result of
\citeauthor{vertes2010}~\cite{vertes2010} that intermediate clustering
maximises the diversity and stability of delay-locked firing sequences.

An important caveat is that the topology sweep was performed at
$N\!=\!500$ on single parametric runs.
Ensemble statistics characterising phase-boundary variance,
network-size effects, and the $N$-scaling of \PNG yield at the
small-world transition constitute the immediate next experimental
steps.

\subsection{Recurrence Plots as universal, label-free \PNG decoders}
\label{sec:disc_rp}

The \RP framework addresses a long-standing methodological limitation:
existing detection approaches require either explicit enumeration of
participating neurons~\cite{martinezmoisy2009} or real-time
simulation instrumentation~\cite{chrolcannon2017}.
The \RP approach requires only the binary spike matrix and operates
entirely in phase space, with no anatomical prior.

The four hypotheses of Sec.~\ref{sec:rp_th} provide rigorous
theoretical grounding: \PNGs produce unit-slope diagonals in
$\mathbf{R}$ because their causal chains propagate at the same
temporal velocity across distinct activations.
This signature is invisible in the raster (Fig.~\ref{fig:rp}, top)
but unambiguous in $\mathbf{R}$ (bottom).

The value $\DET\!\approx\!0.65$ deserves careful interpretation.
A purely stochastic AI network would yield $\DET\!\approx\!0$: all
recurrent points would be isolated.
The observed value indicates that ${\sim}65\%$ of recurrent points
belong to diagonal structures a quantitative measure of the
reproducibility of the dynamical trajectory imposed by the active
\PNG library.
This is consistent with the theoretical picture of cortical dynamics
as organised around a finite manifold of reproducible spatiotemporal
attractors, neither purely random nor rigidly
deterministic~\cite{breakspear2017,cocchi2017}.

The \RP approach is in principle extensible to in-vivo multi-electrode
or calcium-imaging data, provided recording duration is sufficient to
observe at least two activations of the target \PNG.

\subsection{Limitations}
\label{sec:disc_lim}

\textit{Anchor heuristic.}
The anchor-triplet restriction may systematically undercount \PNGs
requiring larger trigger sets or inhibitory gating.
Extending to larger anchors without exponential cost will require
the sequence-hashing approach of
\citeauthor{chrolcannon2017}~\cite{chrolcannon2017}.

\textit{Single-run statistics.}
All topology-sweep results derive from single runs; bootstrapped
phase-boundary estimates and $N$-scaling analyses remain to be
performed.

\textit{RQA calibration.}
The $\DET$, RR, and ENTR values reported here are preliminary
single-run estimates; sensitivity to $\varepsilon$ and $W$, and
bootstrapped confidence intervals, remain to be characterised.

\textit{Deterministic dynamics.}
The present network is driven by Poisson noise but otherwise
deterministic.
The robustness of \PNGs under synaptic unreliability, probabilistic
firing, and continuous membrane noise (via stochastic differential
equations~\cite{ma2023,destexhe2012}) is a critical open question
that constitutes the principal planned extension of this work.

\section{Conclusions}
\label{sec:conclusion}

We have presented a systematic study of Polychronous Neuronal Groups
across three complementary analyses.
Ten hours of \STDP-driven simulation of a biologically grounded
$N\!=\!1{,}000$ network yield 1\,545 unique \PNGs whose structural
backbone is revealed by a bimodal synaptic weight distribution as the
signature of causal-pathway selection by \STDP.
A parametric Watts--Strogatz topology sweep establishes clustering
coefficient as the primary structural determinant of \PNG capacity,
with a $>90\%$ yield reduction from ring lattice to random graph, and
identifies the small-world transition zone as the critical regime.
A sparse-dot-product Recurrence Plot framework identifies \PNGs as
unit-slope diagonal structures in the network's phase-space recurrence
matrix, providing the first label-free analytical tool for \PNG
detection validated against a ground-truth algorithmic library;
$\DET\!\approx\!0.65$ quantifies the reproducibility of the
underlying spatiotemporal dynamics.

These results connect the combinatorial theory of polychronization to
the structural biology of cortical networks and provide a concrete
analytical bridge toward detecting \PNG-like motifs in in-vivo
recordings.

\begin{acknowledgments}
A.D.J. acknowledges financial support from CAPES (Grant No.~0001).
The author FFF gratefully acknowledge the financial support by National Institute of Science and Technology in Innovative Research in Health Sciences – from Nanotechnology to Artificial Intelligence (INCT PICS) sponsored by Brazil’s National Council for Scientific and Technological Development (CNPq), grant no. 408417/2024-2 and grant no. , Coordination of Superior Level Staff Improvement (Capes), grant no. 88887.197686/2025-00, and São Paulo Research Foundation (FAPESP), grant no. 2025/26818-7.Also, FFF acknowledge the financial support by FAPESP grant no. 2025/18142-3, CNPq grants no.311989/2025-0
\end{acknowledgments}

\bibliography{references}

\appendix

\section{Brian2 network simulation code}
\label{app:sim}

The following listing reproduces the complete Brian2 Python
implementation of the baseline \SNN described in
Sec.~\ref{sec:net}.

\section{Offline \PNG detection algorithm}
\label{app:det}

The following listing reproduces the complete Python implementation
of the offline adapted-group detection algorithm described in
Sec.~\ref{sec:detect}.

\end{document}